\documentclass[twocolumn,floatfix,prb,amsmath,amssymb,showkeys]{revtex4}
\usepackage{amsmath,amssymb}
\usepackage{epsfig}
\usepackage{graphicx}
\usepackage{dcolumn}
\usepackage{bm}

\begin{document} 

\title{Estimation of protein folding probability from equilibrium simulations}

\author{Francesco Rao}
\author{Giovanni Settanni}
\author{Enrico Guarnera}
\author{Amedeo Caflisch}

\email[corresponding author, tel: +41 44 635 55 21,
fax: +41 44 635 68 62, e-mail: ]{caflisch@bioc.unizh.ch}

\affiliation{Department of Biochemistry, University of Zurich,
             Winterthurerstrasse 190, CH-8057 Zurich, Switzerland\\
             tel: +41 44 635 55 21, fax: +41 44 635 68 62,\\
             e-mail: caflisch@bioc.unizh.ch}

\date{\today}

\begin{abstract}

The assumption that similar structures have similar folding probabilities
($p_{fold}$) leads naturally to a procedure to evaluate $p_{fold}$ for every
snapshot saved along an equilibrium folding-unfolding trajectory of a
structured peptide or protein.  The procedure utilizes a structurally
homogeneous clustering and does not require any additional simulation.  
It can be used to detect multiple folding pathways as shown for a three-stranded antiparallel $\beta$-sheet peptide investigated by implicit solvent molecular dynamics simulations.

\end{abstract}


\maketitle

\section{introduction}

The folding probability ($p_{fold}$) of a protein conformation saved along a
Monte Carlo or molecular dynamics (MD) trajectory is the probability to fold
before unfolding \cite{Du:On}.  It is a useful measure of kinetic distance from
the folded, i.e., functional state, and can be used to validate transition
state ensemble (TSE) structures, which should have $p_{fold}\approx 0.5$.  Such
validation consists of starting a large number of trajectories from putative
TSE structures with varying initial distribution of velocities and counting the
number of those that fold within a "commitment" time which has to be chosen
much longer than the shortest time-scales of conformational fluctuations and
much shorter than the average folding time \cite{Hubner:Commitment}.  The
concept of $p_{fold}$ calculation originates from a method for determining
transmission coefficients, starting from a known transition state
\cite{Chandler:Statistical} and the identification of simpler transition states
in protein dynamics (e.g., tyrosine ring flips) \cite{Northrup:Dynamical}.  The
approach has been used to identify the otherwise very elusive folding TSE by
atomistic Monte Carlo off-lattice simulations of small proteins with a $G\bar
o$ potential \cite{Li:Constr,Hubner:Commitment}, as well as implicit solvent MD
\cite{Gsponer:Molecular,Rao:The} and Monte Carlo \cite{Lenz:Folding}
simulations with a physico-chemical based potential. The number of trial simulations needed for the reliable evaluation of $p_{fold}$ makes the estimation of the folding probability computationally very expensive.
For this reason, here we propose a method to estimate folding probabilities 
for \textit{all} structures visited in an equilibrium folding-unfolding trajectory without 
any additional simulation. 

\section{Methods}

\begingroup
\squeezetable
\begin{table}[h]
\caption{\label{list}{\sc DRMS} clusters used for the calculation of $P_{f}$.}
\begin{ruledtabular}
\begin{tabular}{ccccccc}
Cluster & 
$P_{f}^C$ \footnote{Cluster-$p_{fold}$ [$P_f^C$, Eq.\ \ref{cluster-pfold}].} & 
$P_{f}$ \footnote{Traditional, i.e., computationally expensive $P_f$ value [Eq.\ \ref{average-pfi}].} & 
$\sigma_{p_{fold}}$ \footnote{Standard deviation of $p_{fold}$ in a cluster [Eq.\ \ref{sigma-pfold}].} & 
$N$ \footnote{Total number of trials used to evaluate $P_f$. For every structure $n_{t}=10$ trials were performed \hbox{($N=n_{t}\ W_{sample}$)}
except for clusters 7 and 25 for which 20 and 50 trials were performed, respectively.} & 
$W$ \footnote{Number of snapshots in the cluster.} & 
$W_{sample}$ \footnote{Number of snapshots used to evaluate $P_f$.
The $W_{sample}$ subset was obtained by selecting structures in a cluster every $|W/W_{sample}|$ saved conformations.} \\
\hline
\hline
     1 &   0.00 &   0.03 &   0.04 &  150 &  144 &   15\\
     2 &   0.11 &   0.05 &   0.06 &  150 &  449 &   15\\
     3 &   0.06 &   0.05 &   0.07 &  120 &   36 &   12\\
     4 &   0.08 &   0.07 &   0.08 &  140 &  555 &   14\\
     5 &   0.10 &   0.08 &   0.06 &  100 &   10 &   10\\
     6 &   0.13 &   0.12 &   0.18 &  160 &  911 &   16\\
     7 &   0.25 &   0.16 &   0.07 &   80 &    4 &    4\\
     8 &   0.23 &   0.20 &   0.31 &  150 &  141 &   15\\
     9 &   0.21 &   0.22 &   0.15 &  140 &  178 &   14\\
    10 &   0.12 &   0.23 &   0.20 &  120 &   48 &   12\\
    11 &   0.57 &   0.25 &   0.14 &  140 &   14 &   14\\
    12 &   0.05 &   0.27 &   0.19 &  100 &   19 &   10\\
    13 &   0.23 &   0.29 &   0.38 &  140 &  391 &   14\\
    14 &   0.08 &   0.30 &   0.15 &  120 &   12 &   12\\
    15 &   0.72 &   0.35 &   0.23 &  130 &  129 &   13\\
    16 &   0.19 &   0.38 &   0.18 &  130 &   26 &   13\\
    17 &   0.38 &   0.44 &   0.39 &  160 &   16 &   16\\
    18 &   0.38 &   0.51 &   0.28 &  160 &   16 &   16\\
    19 &   0.65 &   0.60 &   0.29 &  100 &   20 &   10\\
    20 &   0.57 &   0.61 &   0.35 &   70 &    7 &    7\\
    21 &   0.48 &   0.63 &   0.32 &  140 &   27 &   14\\
    22 &   0.74 &   0.65 &   0.40 &  140 &  539 &   14\\
    23 &   0.68 &   0.66 &   0.18 &  140 &   28 &   14\\
    24 &   0.38 &   0.71 &   0.24 &  130 &   13 &   13\\
    25 &   0.50 &   0.72 &   0.20 &  100 &    2 &    2\\
    26 &   0.82 &   0.76 &   0.31 &  170 &   17 &   17\\
    27 &   0.50 &   0.78 &   0.14 &  120 &   12 &   12\\
    28 &   0.78 &   0.78 &   0.22 &  180 &   18 &   18\\
    29 &   0.70 &   0.79 &   0.19 &  130 &  189 &   13\\
    30 &   0.77 &   0.79 &   0.17 &  150 &   30 &   15\\
    31 &   0.85 &   0.81 &   0.11 &  130 &   13 &   13\\
    32 &   0.91 &   0.83 &   0.20 &  140 &  401 &   14\\
    33 &   0.90 &   0.85 &   0.27 &  100 &   20 &   10\\
    34 &   0.85 &   0.85 &   0.10 &  120 &   48 &   12\\
    35 &   0.94 &   0.88 &   0.13 &  170 & 1990 &   17\\
    36 &   0.71 &   0.94 &   0.07 &   70 &    7 &    7\\
    37 &   0.95 &   0.95 &   0.06 &  150 &  855 &   15\\
\end{tabular}
\end{ruledtabular}
\end{table}
\endgroup

\subsection{Molecular dynamics simulations}

Beta3s is a designed 20-residue sequence whose
solution conformation has been investigated by NMR spectroscopy
\cite{DeAlba:Denovo}. The NMR data indicate that beta3s in aqueous solution
forms a monomeric (up to more than 1mM concentration) triple-stranded
antiparallel $\beta$-sheet, in equilibrium with the
denatured state \cite{DeAlba:Denovo}. We have previously shown that in
implicit solvent \cite{Ferrara:Evaluation} molecular dynamics simulations
beta3s folds reversibly to the NMR solution conformation, irrespective of the
starting structure \cite{Ferrara:Folding}.  
Recently, four molecular dynamics simulations of beta3s were performed at 330 K
for a total simulation time of 12.6 $\mu$s \cite{Cavalli:Fast}.  There are 72
folding events and 73 unfolding events and the average time required to go from
the denatured state to the folded conformation is 83 ns.  The 12.6 $\mu$s of
simulation length is about two orders of magnitude longer than the average
folding or unfolding time, which are similar because at 330 K the native and
denatured states are almost equally populated \cite{Cavalli:Fast}. For the
$p_{fold}$ analysis the first 0.65 $\mu$s of each of the four simulations were
neglected so that along the 10 $\mu$s of simulations there are a total of 500000 snapshots because coordinates were saved every 20 ps.

The simulations  were performed with
the program CHARMM {\cite{Brooks:CHARMM}}.  Beta3s was modeled by explicitly
considering all heavy atoms and the hydrogen atoms bound to nitrogen or oxygen
atoms (PARAM19 force field {\cite{Brooks:CHARMM}}).  A mean field approximation
based on the solvent accessible surface was used to describe the main effects
of the aqueous solvent on the solute \cite{Ferrara:Evaluation}.  The two
surface tension-like parameters of the solvation model were optimized without using beta3s.  The
same force field and implicit solvent model have been used recently in
molecular dynamics simulations of the early steps of ordered aggregation
\cite{Gsponer:Therole}, and folding of structured peptides\cite{Ferrara:Evaluation,Ferrara:Folding}, as well as small
proteins of about 60 residues \cite{Gsponer:Role}. Despite
the absence of collisions with water molecules, in the simulations with
implicit solvent the separation of time scales is comparable with that observed
experimentally.  Helices fold in about 1 ns \cite{Ferrara:Thermodynamics},
$\beta$-hairpins in about 10 ns \cite{Ferrara:Thermodynamics} and
triple-stranded $\beta$-sheets in about 100 ns \cite{Cavalli:Fast}, while the
experimental values are $\sim$0.1 $\mu$s \cite{Eaton:Fast}, $\sim$1 $\mu$s
\cite{Eaton:Fast} and $\sim$10 $\mu$s \cite{DeAlba:Denovo}, respectively.

\subsection{Clusterization}

\begin{figure}[h]
\includegraphics[angle=-90,width=8cm]{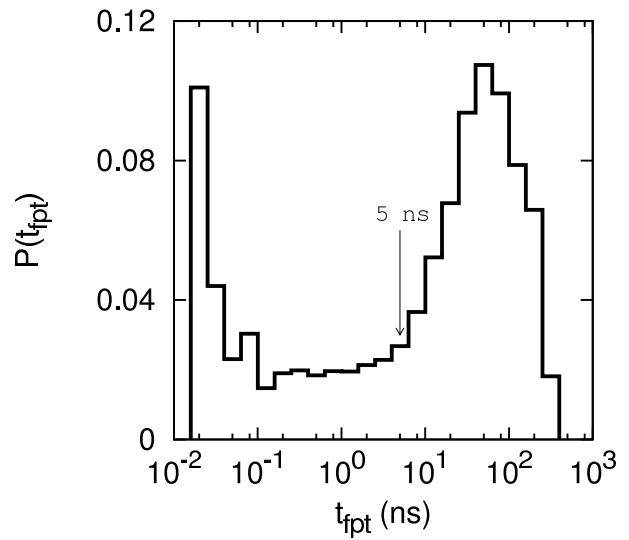}
\caption{Probability distribution for the first passage time (fpt) to the most
populated cluster (\emph{folded state}) of the DRMS 1.2 \AA\
clusterization.}
\label{fig.pfpt}
\end{figure}

The 500000 conformations obtained from the simulations of beta3s (see above) were clustered by the leader algorithm \cite{hartigan}. Briefly, the first structure defines the first cluster and each
subsequent structure is compared with the set of clusters found so far until the first similar structure is found.
If the structural deviation (see below) from the first conformation of all of the known clusters exceeds a given threshold, a new cluster is defined.
The leader algorithm is very fast even when analyzing large sets of
structures like in the present work.
The results presented here were obtained with a structural comparison based on the Distance Root Mean Square (DRMS) deviation considering all distances involving C$_\alpha$ and/or C$_\beta$ atoms and a cutoff of 1.2 \AA.  This yielded 78183 clusters.
The DRMS and root mean square deviation of atomic coordinates (upon
optimal superposition) have been shown to be highly correlated \cite{Hubner:Commitment}. The DRMS cutoff of 1.2~\AA\ was chosen on the basis of the distribution of the pairwise DRMS values in a subsample of the wild-type trajectories.  The distribution shows
two main peaks that originate from intra- and inter-cluster distances,
respectively
(data not shown).  The cutoff is located at the minimum between the two
peaks.
The main findings of this work are valid
also for clusterization based on secondary structure similarity
\cite{Rao:The}
(see Suppl.\ Mat.).



\subsection{Folding probability}

\begin{figure*}
\includegraphics[width=7.0cm, angle=-90]{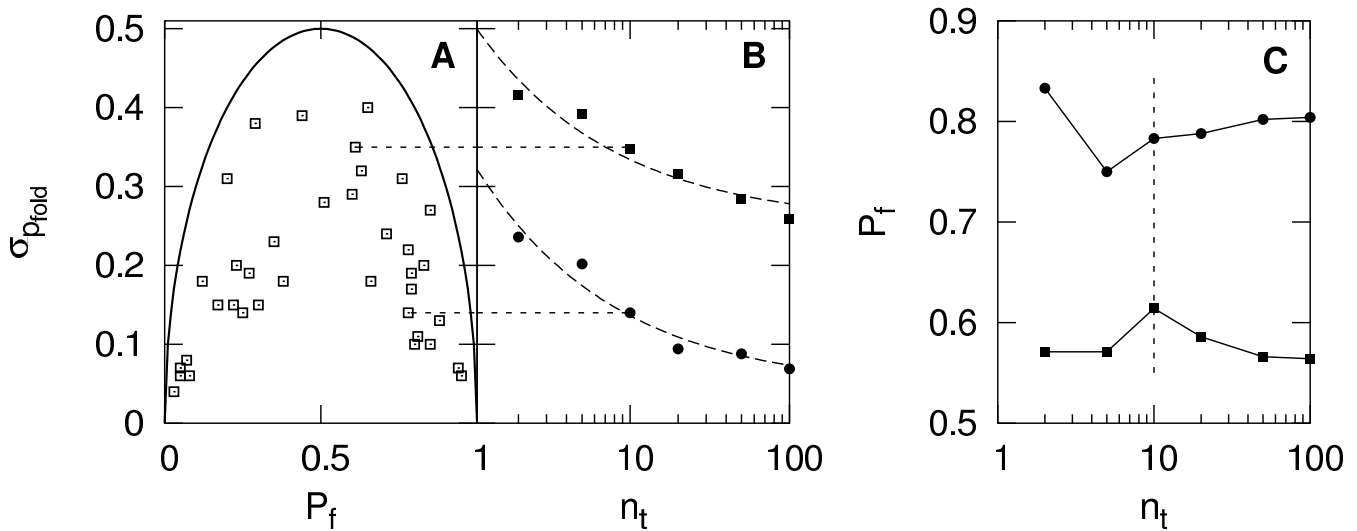}
\caption{Standard deviation 
$\sigma_{p_{fold}}=\sqrt{\left< (p_{fold}(i)-P_{f}[\alpha])^2
\right>_{i \in \alpha}}$ of the $p_{fold}$ for the 37 DRMS clusters used in the study.
\textbf{(A)} $\sigma_{p_{fold}}$ as a function of $P_{f}$ compared to a
Bernoulli distribution (solid line). Ten trials were performed for each
snapshot.  The largest values for the standard deviation are located around the
0.5 region and this is probably due to the Bernoulli process ($\theta=0,1$)
used for the calculation of $p_{fold}$. \textbf{(B)} $\sigma_{p_{fold}}$
dependence on the number of trials used to evaluate $p_{fold}$. The dashed curves are fits with a $\frac{a}{\sqrt{x}}+b$ function.  The horizontal
dashed lines are drawn to help identifying in \textbf{A} the two clusters used
in \textbf{B}.   \textbf{(C)} Dependence of $P_f$  on the number of trials $n_t$ for the two clusters used in \textbf{B}.}
\label{fig.sigmapfold}
\end{figure*}

For the computation of $p_{fold}$ a criterion ($\Phi$) is needed to determine
when the system reaches the folded state.
Given a clusterization of the structures, a natural choice for $\Phi$ is the visit of the most populated cluster which for
structured peptides and proteins is not degenerate (other criteria are also possible, e.g., fraction of native contacts $Q$ larger than a given threshold).
Given $\Phi$ and a commitment time ($\tau_{commit}$), 
the folding probability $p_{fold}(i)$ of an MD snapshot $i$ is computed as \cite{Du:On,Hubner:Commitment}
\begin{equation} 
p_{fold}(i)=\frac{n_{f}(i)}{n_t(i)}\ 
\label{pfi}
\end{equation}
where $n_{f}(i)$ and $n_t(i)$ are the number of trials started from snapshot $i$
which reach within a time $\tau_{commit}$ the folded state and the total number of trials, respectively.  

Every simulation started from snapshot $i$ can 
be considered as a Bernoulli trial of a random variable
$\theta$ with value 1 (folding within $\tau_{commit}$) or 0 
(no folding within $\tau_{commit}$).  The variable
$\theta$ has average and variance on the average of the form:
\begin{equation}
\begin{split} 
\langle \theta \rangle = &
p_{fold}\ =  \frac{1}{n_t}\sum_{i=1}^{n_t} \theta_i \\
\sigma^2_{\left< \theta \right>} = & \frac{1}{n_t}p_{fold}(1-p_{fold})
\end{split}
\end{equation}
where $n_t$ is the total number of trials and the accuracy on the $p_{fold}$
value increases with $n_t$.

In Fig.\ \ref{fig.pfpt} the distribution of the first passage time (fpt)
to the folded state is shown. The double
peak shape of the distribution provides evidence for the different time scales
between \emph{intra}-basin and \emph{inter}-basin transitions.
A value of 5 ns is chosen
for $\tau_{commit}$ because events with
smaller time scales correspond to the diffusion within the native free-energy
basin, while events with larger time scales are transitions from other basins
to the native one, i.e., folding/unfolding events \cite{Cavalli:Fast}.

\section{Folding probability from equilibrium trajectories}

\begin{figure*}[t]
\includegraphics[width=11.5cm, angle=-90]{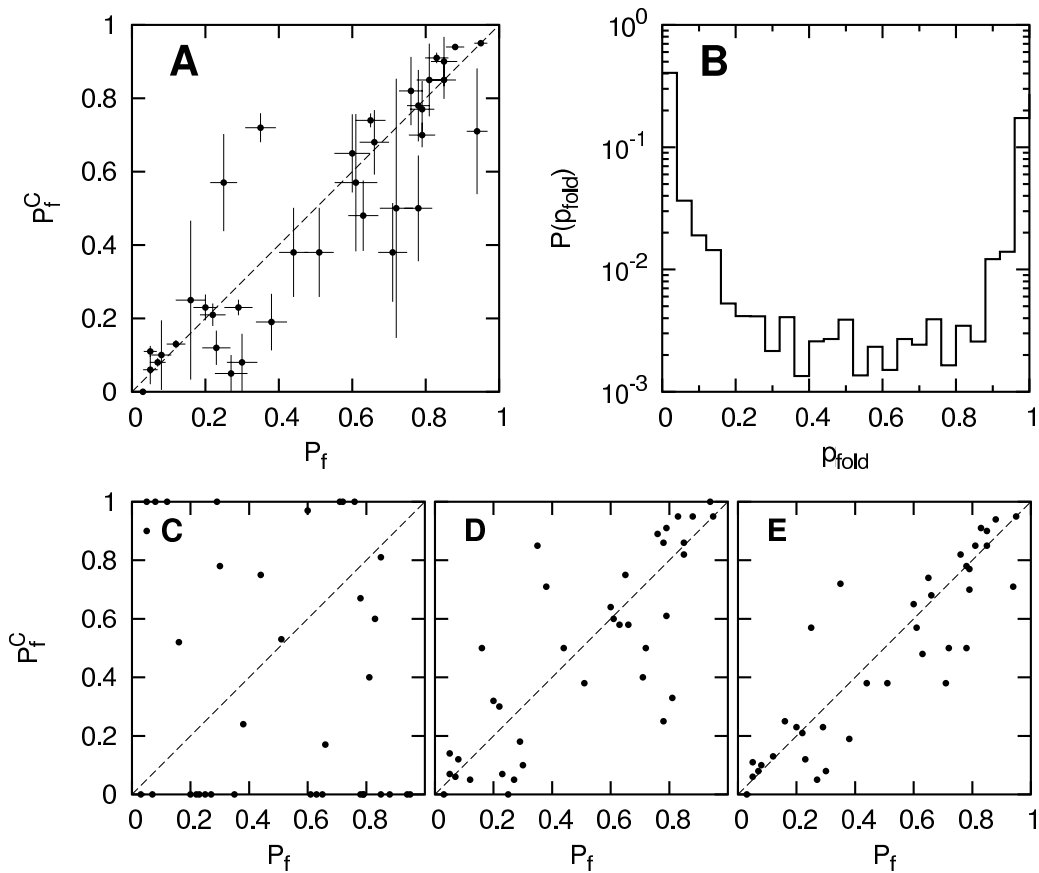}
\caption{Cluster folding probability $P_{f}^{C}$. \textbf{(A)} Scatter plot of $P_{f}^C$ versus $P_{f}$.
The DRMS 1.2 \AA\ clusterization and the folding criterion $\Phi$ 
(reaching the most populated cluster within $\tau_{commit}=5$ ns)
were used. \textbf{(B)} Probability distribution of the $p_{fold}$ value for the 500000 snapshots saved along the $10\ \mu s$ MD trajectory. The folding probability for snapshot $i$ is computed as $p_{fold}(i)=P_f^C[\alpha]$ for $i \in \alpha$. \textbf{(C-E)} Scatter plot of $P_{f}^C$ versus $P_{f}$ for 1.0, 5.0, and 10 $\mu s$ of simulation time, respectively.}
\label{fig.scatterpfold}
\end{figure*}

The basic assumption of the present work is that conformations that
are structurally similar have the same kinetic behavior, hence they have
similar values of $p_{fold}$.  Note that the opposite is not necessarily true as explained in Section IV for the TSE and the denatured state. To exploit this assumption, snapshots saved along
a trajectory are grouped in structurally similar clusters\cite{Symbolic}.
Then, the $\tau_{commit}$-segment of MD trajectory following each snapshot is
analyzed to check if the folding condition $\Phi$ is met (i.e, the snapshot
"folds").  For each cluster, the ratio between the snapshots which lead to
folding and the total number of snapshots in the cluster is defined as the
cluster-$p_{fold}$ ($P_{f}^C$; throughout the text uppercase $P$ and lowercase
$p$ refer to folding probability for clusters and individual snapshots,
respectively).  This value is an approximation of the $p_{fold}$ of any single
structure in the cluster which is valid if the cluster consists of structurally
similar conformations.  In other words, the occurrence of the folding event for
the snapshots of a given cluster can be considered as a Bernoulli trial of a
random variable $\theta$.  The average of $\theta$ and variance on the average
for the set of snapshots belonging to a given cluster $\alpha$ can be written
as: 
\begin{equation}
\begin{split} 
P_{f}^C [\alpha] & = \langle \theta \rangle =  
\frac{1}{W}\sum_{i=1}^{W} \theta_i\ , \qquad i \in \alpha \\
\sigma^2_{\left< \theta \right>} & = \frac{1}{W}P_{f}^C(1-P_{f}^C)
\end{split} 
\label{cluster-pfold}
\end{equation}
where $W$ is the number of snapshots in cluster $\alpha$.  $P_{f}^C$ is
the average folding probability over a set of structurally homogeneous
conformations. Using the clustering and the folding criterion $\Phi$
introduced above, values of $P_{f}^C$ for the 78183 clusters can be computed
by Eq.\ \ref{cluster-pfold}, i.e., the number of conformations of the
cluster that fold within 5 ns divided by the total number of conformations
belonging to the cluster.

In this article we provide evidence that the basic assumption mentioned above,
that is, similar conformations have similar folding probabilities, holds in
the case of beta3s, a three-stranded antiparallel $\beta-$sheet peptide investigated by MD \cite{Cavalli:Fast}.
Moreover,  we show that the computationally expensive
\begin{equation} 
P_{f}[\alpha] = \frac{1}{W}\sum_{i=1}^{W} p_{fold}(i)\ ,
\qquad i \in \alpha
\label{average-pfi}
\end{equation}
which is measured by 
starting several simulations from each snapshot $i$ in the cluster $\alpha$
with $W$ snapshots, is well approximated by $P_{f}^C$ whose evaluation is
straightforward.

To test the assumption that similar structures have similar $p_{fold}$ and to
compare the values of $P_{f}^C$ with those obtained from the standard approach
\cite{Du:On}, folding probabilities $P_{f}$ were computed for the structures of
37 clusters by starting several 5 ns MD runs from each structure and counting
those that fold (Eq.~\ref{pfi} and~\ref{average-pfi}).  The 37 clusters chosen
among the 78183 include both high- and low-populated clusters with $P_{f}^C$
values evenly distributed in the range between 0 and 1 (see Tab.\ 1).  In the
case of large clusters a subset of snapshots is considered for the
computation of $P_{f}$. In those cases $W$ is replaced in Eq.~\ref{average-pfi} by $W_{sample}<W$ that is the number of snapshots involved
in the calculation.

\begin{figure*}[p]
\includegraphics[width=12cm]{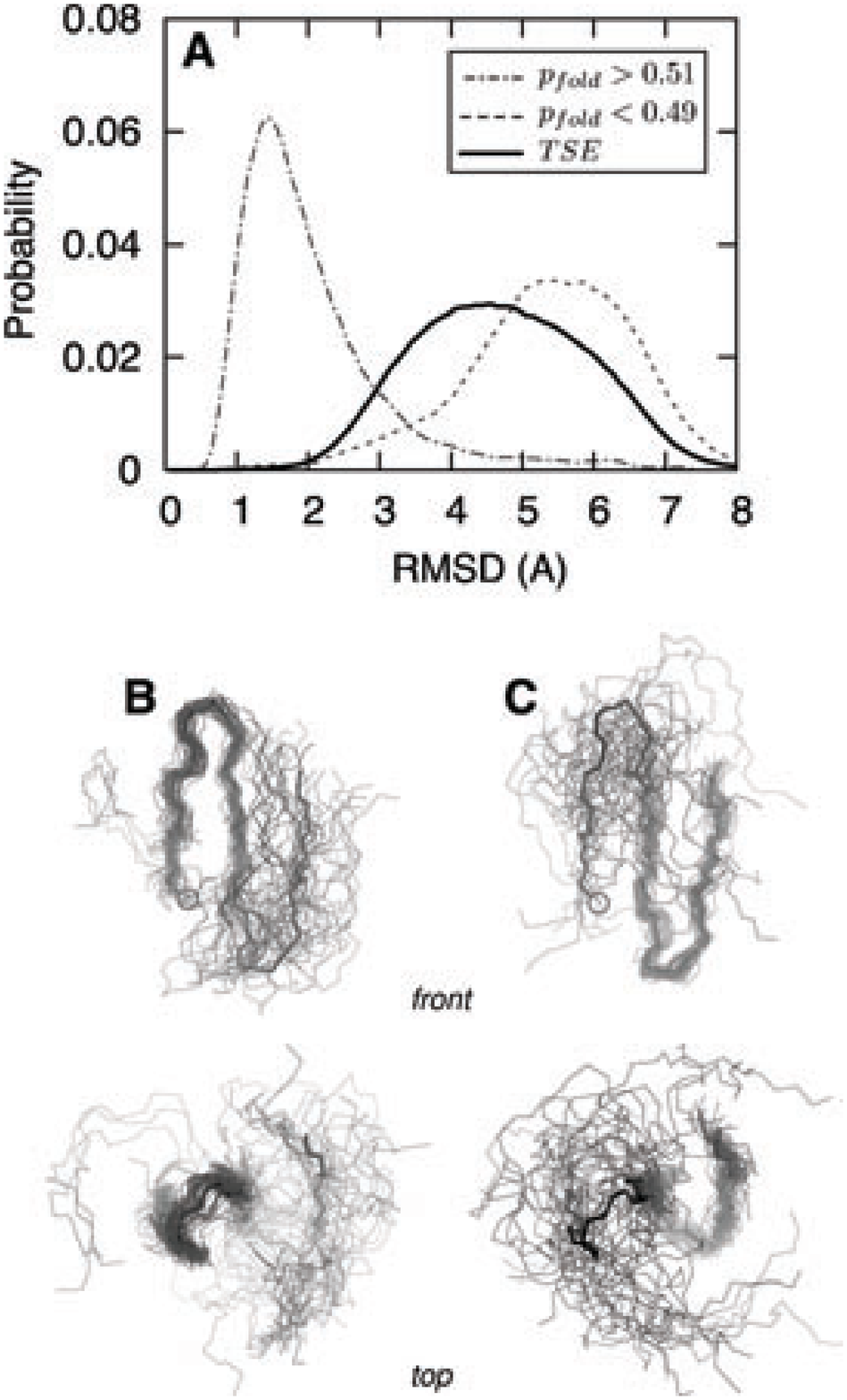}
\caption{Transition state ensemble (TSE) of beta3s. \textbf{(A)} RMSD pairwise distribution for structures with $p_{fold}>0.51$ (native state), $0.49< p_{fold}< 0.51$ (TSE), and $p_{fold}<0.49$ (denatured state). \textbf{(B)} Type I and \textbf{(C)} type II transition states (thin lines). Structures are superimposed on residues 2-11 and 10-19 with an average pairwise RMSD of 0.81 and 0.82 \AA\ for type I and type II, respectively. For comparison, the native state is shown as a thick line with a circle to label the N-terminus.}
\label{fig.pair}
\end{figure*}

The standard deviation of $p_{fold}$ in a cluster is computed as
\begin{equation}
\sigma_{p_{fold}}=\sqrt{\left< (p_{fold}(i)-P_{f}[\alpha])^2 \right>_{i \in \alpha}} 
\label{sigma-pfold}
\end{equation}
In the case of full kinetic inhomogeneity, i.e., random grouping of snapshots,
the $p_{fold}$ value  for all snapshots in a given cluster will be equal to 0
or 1, indicating the coexistence (in the same cluster) of structures that
either exclusively fold or unfold. In this case $\sigma_{p_{fold}}$ reflects
the Bernoulli distribution (see Supp. Mat.).  Fig.~\ref{fig.sigmapfold}A shows that, even when only $n_t=10$ runs
per snapshot are used to compute $p_{fold}$, $\sigma_{p_{fold}}$ values are not
compatible with those of a Bernoulli distribution.  Moreover the values of the
standard deviation decrease when the number of trials $n_t$ increases,
as reported in Fig.~\ref{fig.sigmapfold}B for two sample clusters. The asymptotic value
of $\sigma_{p_{fold}}$  ($n_t \rightarrow \infty$) for these two data sets is
of 0.05 and 0.2. This value cannot reach zero because snapshots in a cluster
are similar but not identical.  These results suggest that snapshots inside the same
cluster are kinetically homogeneous and a statistical description of $p_{fold}$ can
be adopted, that is, folding probabilities are computed as cluster averages
(instead of single snapshots) by means of $P_f$ and $P_f^C$.

We still have to verify that $P_{f}^C$ indeed approximates the computationally expensive $P_{f}$. Namely, for the 37 clusters mentioned above a correlation of 0.89 between $P_{f}^C$ and $P_{f}$ is found with a slope of 0.86 (see Fig.~\ref{fig.scatterpfold}A and
Tab.~1), indicating that the procedure is able to estimate folding
probabilities for clusters on the folding-transition barrier ($P_{f}\sim 0.5$)
as well as in the folding ($P_{f}\sim 1.0$) or unfolding ($P_{f}\sim 0.0$)
regions.  The error bars for $P_{f}^C$ in Fig.~\ref{fig.scatterpfold}A are derived from the
definition of variance given in Eq.\ \ref{cluster-pfold}.  In the same spirit
of Eq.\ \ref{cluster-pfold} the folding probability $P_{f}$ and its variance
are written as
\begin{equation} 
\begin{split}
P_{f} = & \left < \theta \right> = \frac{1}{N}\sum_{i=1}^N \theta_i \\
\sigma^2_{\left< \theta \right>} = & \frac{1}{N} P_{f}(1-P_{f})
\end{split}
\label{sigmapfc} 
\end{equation}
where $N=\sum n_t$ is the total number of runs and $\theta$ is equal to 1 or 0,
if the run folded or unfolded, respectively.  Note that the same number of runs
$n_t$ has been used for every snapshot of a cluster. The large vertical error bars in
Fig.~\ref{fig.scatterpfold}A correspond to clusters with less than 10 snapshots.  The largest
deviations between $P_{f}$ and $P_{f}^C$ are around the $0.5$ region. This is
due to the limited number of crossings of the folding barrier observed in the
MD simulation (Fig.~\ref{fig.scatterpfold}B, around 70 events of folding \cite{Cavalli:Fast}).
Improvements in the accuracy for the estimation of $P_f$ are achieved as the
number of folding events, i.e., the simulation time, increases (Fig.~\ref{fig.scatterpfold}C-E).  

The two main results of this study, i.e., the kinetic homogeneity of the clusters and the
validity of $P_f^C$ as an approximation of $P_f$, are robust with respect to the choice of the clusterization.  Similar results
can be obtained also with different flavors of conformation space partitioning,
as long as they group together structurally homogeneous conformations, e.g., clusterization based on root mean square deviation of atomic coordinates (RMSD) or secondary structure strings (see Supp.\ Mat.).
The latter are appropriate for structured peptides but not for proteins with irregular secondary structure because of string degeneracy. Note that partitions
based on order parameters (like native contacts) are usually unsatisfactory and
not robust. This is mainly due to the fact that clusters defined in this way
are characterized by large structural heterogeneities \cite{Rao:The}.

\section{Analysis of transition state ensemble}

The folding probability of structure $i$ is estimated as $p_{fold}(i)=P_{f}^{C}[\alpha]$ for $i \in \alpha$. This approximation allows to plot the pairwise RMSD distribution of beta3s structures with 
$p_{fold}>0.51$ (native state), $0.49< p_{fold}< 0.51$ (transition state ensemble, TSE), and $p_{fold}<0.49$ (denatured state) (Fig.~\ref{fig.pair}A). For the native state, the distribution is peaked around low values of RMSD ($\sim 1.5$~\AA) indicating that structures with $p_{fold}>0.51$ are structurally similar and belong to a non-degenerate state. The statistical weight of this group of structures is 49.4\% and corresponds to the expected statistics for the native state because the simulations are performed at the melting temperature. 
In the case of TSE, the distribution is broad because of the coexistence of heterogeneous structures. This scenario is compatible with the presence of multiple folding pathways. Beta3s folding was already shown to involve two main average pathways depending on the sequence of formation of the two hairpins\cite{Ferrara:Folding,Rao:The}. Here, a \textit{naive} approach based on the number of native contacts\cite{Ferrara:Folding} is used to structurally characterize the folding barrier. TSE structures with number of native contacts of the first hairpin greater than the ones of the second hairpin are called type I conformations (Fig.~\ref{fig.pair}B), otherwise they are called type II (Fig.~\ref{fig.pair}C). In both cases the transition state is characterized by the presence of one of the two native hairpins formed while the rest of the peptide is mainly unstructured. These findings are also in agreement with 
the complex network  analysis of beta3s reported in Ref \onlinecite{Rao:The}. Finally, the denatured state shows a broad pairwise RMSD distribution around even larger values of RMSD  ($\sim 5.5$ \AA), indicating the presence of highly heterogeneous conformations.

\section{Conclusions}

Two main results have emerged from the present study. First, snapshots grouped in structurally homogeneous clusters are
characterized by similar values of $p_{fold}$. This result justifies the use of
a statistical approach for the study of the kinetic properties of the
structures sampled along a simulation.  Second, given a set of structurally
homogeneous clusters and a folding criterion, it is possible to obtain a first
approximation of the folding probability for every structure sampled along an
equilibrium folding-unfolding simulation. Thus, the cluster-$p_{fold}$ is a quantitative
measure of the kinetic distance from the native state and is computationally very cheap\cite{Computation}. Furthermore, it can be used to detect multiple folding pathways. The accuracy in the identification of the transition state ensemble improves as the number of folding events observed in the simulation increases.
%
%
Recently the cluster-$p_{fold}$ approach has been used to identify the transition state ensemble of a large set of beta3s mutants (for a total of 0.65~$ms$ of simulation time\cite{Settanni:Phi}), which would have been impossible with traditional methods.  As a further
application, the cluster-$p_{fold}$ procedure can be used to validate TSE conformations
obtained by wide-spread $G\bar o$ models.

\begin{acknowledgments}
We thank Stefanie Muff for useful and stimulating discussions and comments to the manuscript.
We also thank Dr.\ Emanuele Paci for interesting discussions.  
We acknowledge an anonymous referee for suggesting the use of cluster-$p_{fold}$ to detect multiple pathways.
The molecular 
dynamics simulations were performed on the Matterhorn Beowulf cluster at the
Informatikdienste of the University of Zurich.  We thank C.\ Bollinger, Dr.\
T.\ Steenbock, and Dr.\ A.\ Godknecht for setting up and maintaining the
cluster.  This work was supported by the Swiss National Science Foundation
grant nr. 205321-105946/1.
\end{acknowledgments}

\bibliography{a-bib}

\clearpage

\onecolumngrid

\markright{\centerline{\LARGE\sc Supplementary Material}}
\pagestyle{myheadings}

\clearpage

\setcounter{section}{0}

\setcounter{page}{1}
\renewcommand{\thepage}{S-\arabic{page}}

\setcounter{figure}{0}
\renewcommand{\thefigure}{S\arabic{figure}}

\setcounter{table}{0}
\renewcommand{\thetable}{S-\Roman{table}}

\section{Secondary structure clusterization}

Recently, the secondary structure has been used to cluster the conformation space
of peptides (F. Rao et al, JMB 342, 299, 2004). Secondary structure along an MD
simulation trajectory can be easily calculated using known algorithms
(C.A.F. Andersen et al, Structure 10, 174, 2002).  A cluster is a single string
of secondary structure, e.g., the most populated conformation for beta3s is
{\tt -EEEESSEEEEEESSEEEE-} where "{\tt E}", "{\tt S}", and "{\tt -}" stand for
extended, turn, and unstructured, respectively.  There are 8 possible "letters"
in the secondary structure "alphabet": "{\tt H}", "{\tt G}", "{\tt I}", "{\tt
E}", "{\tt B}", "{\tt T}", "{\tt S}", and "{\tt -}", standing for $\alpha$
helix, 3/10 helix, $\pi$ helix, extended, isolated $\beta$-bridge, hydrogen
bonded turn, bend, and unstructured, respectively.  Since the N- and C-terminal
residues are always assigned an "{\tt -}" a 20-residue peptide can in principle
assume $8^{18}\simeq 10^{16}$ conformations.

\begin{figure*}[h]
\includegraphics[angle=-90,width=70mm]  {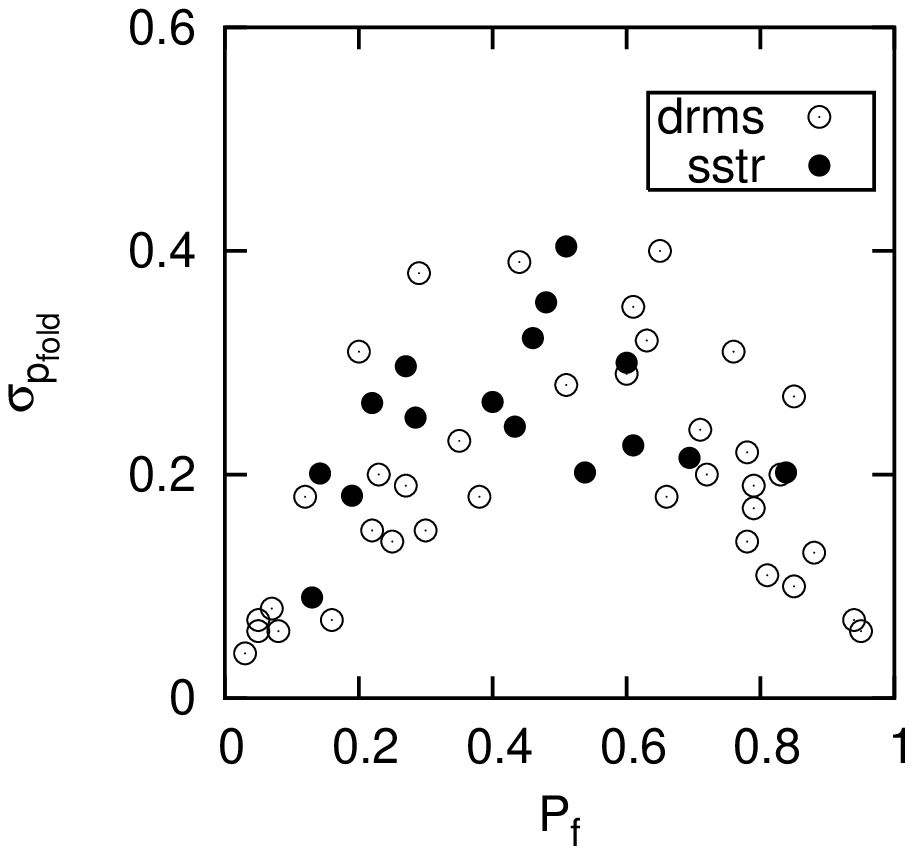}
\includegraphics[angle=-90,width=80mm]  {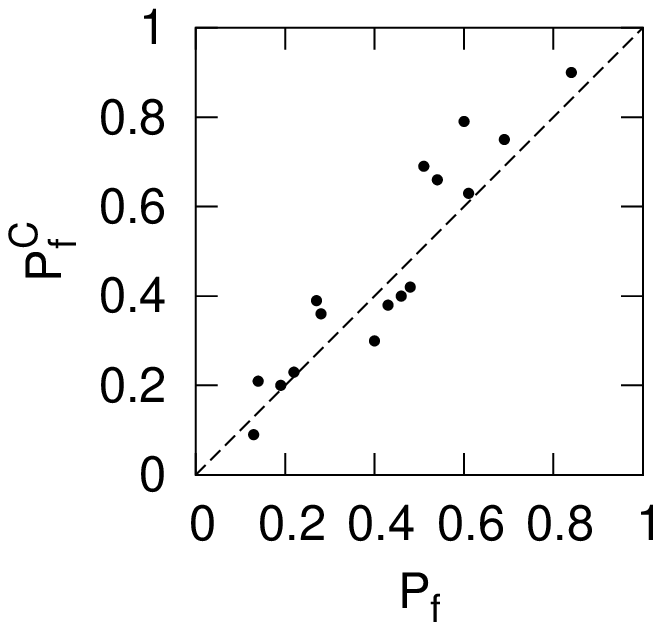}
\caption{\textbf{(left)} $p_{fold}$ standard deviation inside a cluster for 16
secondary structure (\emph{sstr}) and 37 DRMS 1.2 \AA\ clusters.  Both
\emph{sstr} and DRMS 1.2 \AA\ clusterizations are defined by similar
fluctuations.  \textbf{(right)} Scatter plot of $P_{f}^C$ versus $P_{f}$ for
\emph{sstr} clusterization.  In this case the folding criteria used is based on
the native contacts $Q$ (Settanni et al., PNAS 102, 628, 2005). A folding
(unfolding) event is realized when $Q>0.85$ ($Q<0.15$).}
\end{figure*}

\begin{figure*}[h]
\includegraphics[angle=-90,width=90mm]  {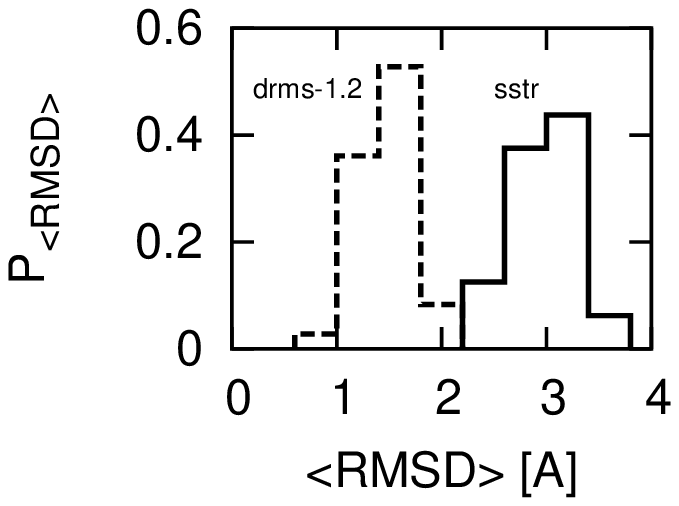}
\includegraphics[angle=-90,width=90mm]  {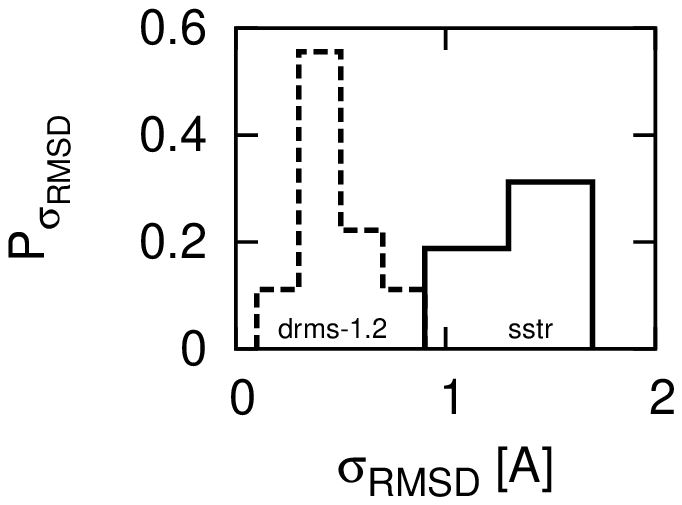} 
\caption{\textbf{(top)} Probability to have a given pairwise
root mean square deviation (RMSD) inside a cluster for the secondary
structure (\emph{sstr}) and DRMS 1.2 \AA\ clusterizations. \textbf{(bottom)}
Probability to have a given variance for the RMSD inside a cluster.
Both plots show that secondary structure clusters are less structurally
homogeneous than DRMS 1.2 \AA\ clusters.}
\end{figure*}

\clearpage

\section{First passage times}

The first passage time (fpt) to a given cluster $\alpha$ is computed as the
time along the MD trajectory that any given snapshot takes to the first
subsequent snapshot belonging to $\alpha$. In fig.\ S3 the fpt distribution 
to the folded state is
shown for two different clusterizations of the conformation space. The double
peak shape of the distribution provides evidence of the different time scales
between \emph{intra}-basin and \emph{inter}-basin transitions. The wider shape
of the \emph{intra}-basin peak for the secondary structure clusterization is
consistent with the higher degree of structural diversity with
respect to the DRMS 1.2 \AA\ clusterization (see previous section).

\begin{figure*}[h]
\includegraphics[angle=-90,width=90mm]  {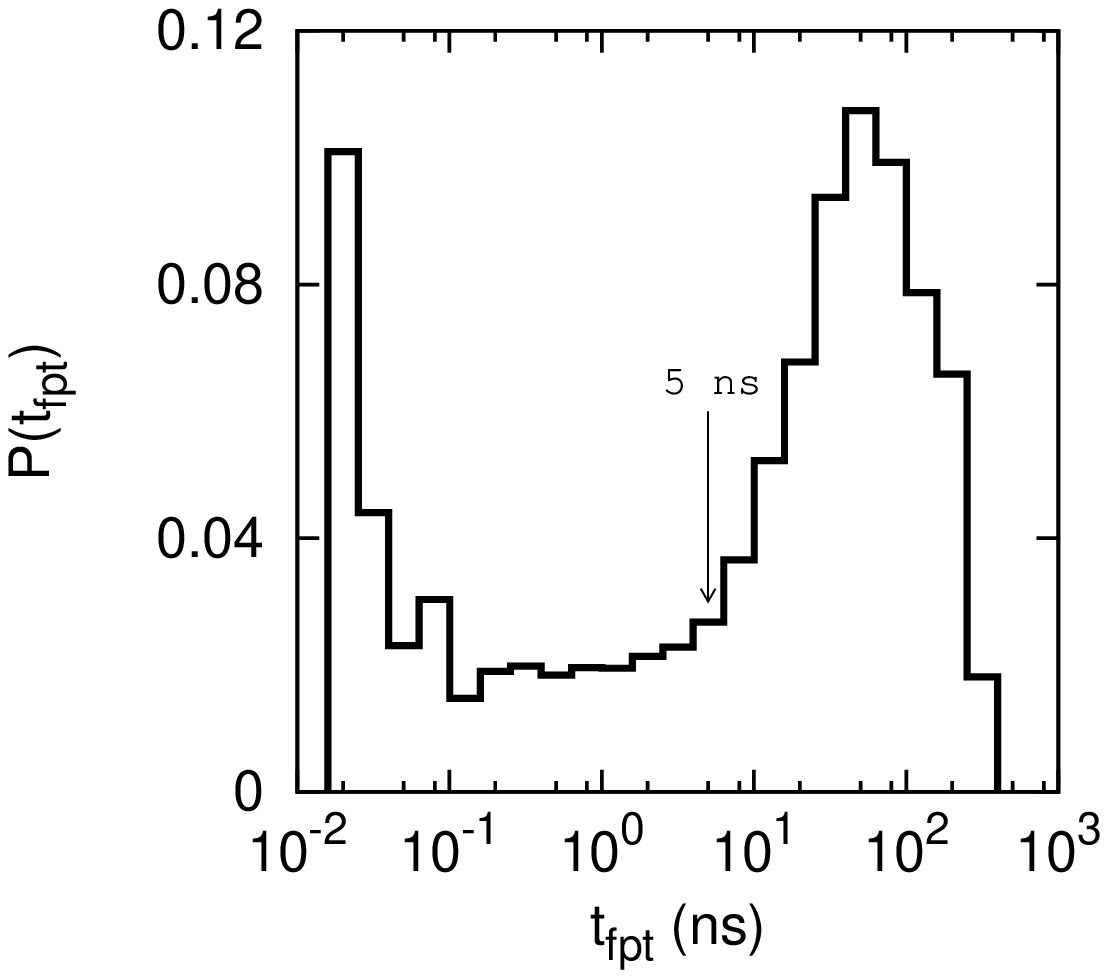}
\includegraphics[angle=-90,width=90mm]  {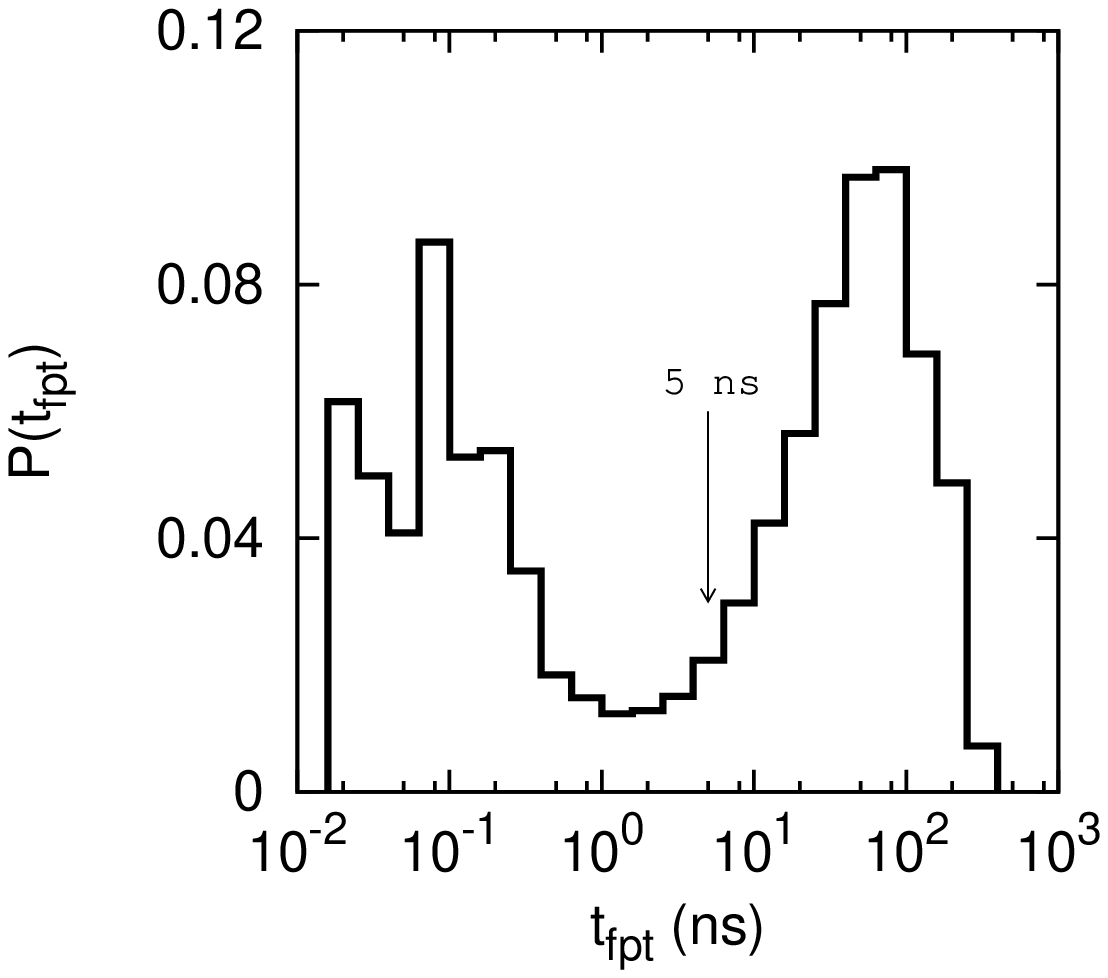}
\caption{Probability distribution for the first passage times (fpt) to the most
populated cluster (\emph{folded state}). \textbf{(top)} DRMS 1.2 \AA\
clusterization.  \textbf{(bottom)} Secondary structure clusterization.}
\end{figure*}

\clearpage

\section{Random clusterization}

\begin{figure*}[h]
\includegraphics[angle=-90,width=80mm]  {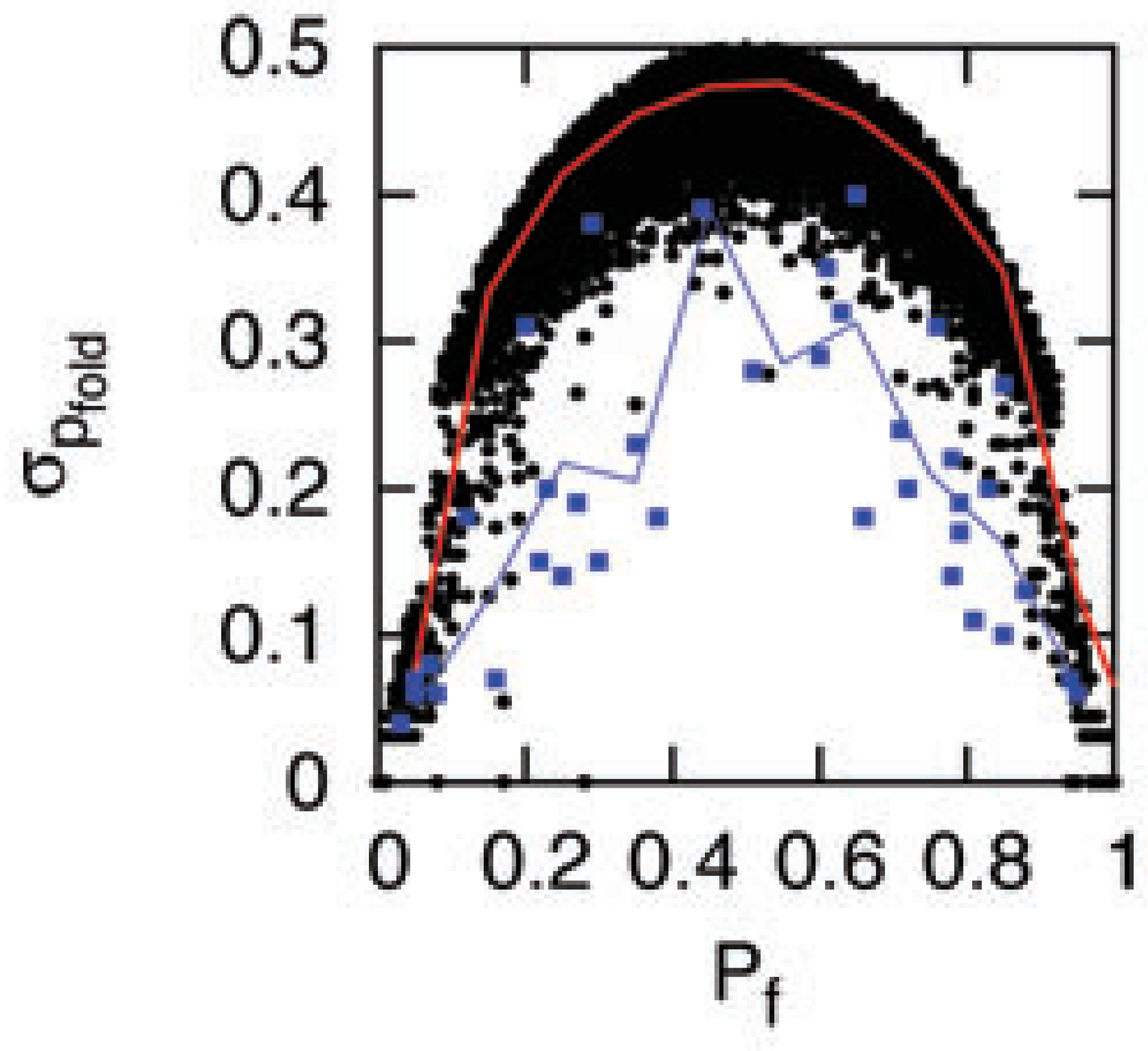}
\caption{Standard deviation $\sigma_{p_{fold}}$ for a random clusterization.
Black dots, red curve, blue squares, blue curve show
$\sigma_{p_{fold}}$ for the random clusters, its histogram,
$\sigma_{p_{fold}}$ for 37 non-random clusters (see text), and its
histogram, respectively.}
\label{sigmapfold-null}
\end{figure*}

The results of this section were obtained using the DRMS 1.2 \AA\
clusterization. 
In the text evidence was provided that the standard deviation of $p_{fold}$
\begin{equation*} \sigma_{p_{fold}}=\sqrt{\left< (p_{fold}(i)-P_{f}[\alpha])^2 \right>_{i \in \alpha}}
\end{equation*} 
is not compatible with the one of a Bernoulli distribution.  This means that
snapshots in a cluster have similar values of $p_{fold}$ and are kinetically homogeneous.  This is
not the case for a random clusterization of the snapshots.  Since it is not
feasible to compute the $p_{fold}$ for every snapshot of a simulation, the
assumption that $p_{fold}$ of snapshot $i$ is equal to the cluster folding
probability $P_f^C$ of its cluster (as computed in the text) is made.  
Then, snapshots are reshuffled in 50000 random clusters.  The folding
probability for a random cluster $\alpha_R$ is computed as $P_f=\left<
p_{fold}\right>_{\alpha_R}$.  Most of the snapshots will have $p_{fold}$ close
to $1$ or $0$ (see Fig.\ 3B in the text) and because of the random grouping,
i.e., no kinetic homogeneity, the above standard deviation $\sigma_{p_{fold}}$
resembles the one of a Bernoulli distribution as shown in Fig.\
\ref{sigmapfold-null}. Data obtained from a DRMS 1.2 \AA\ clusterization
deviates from this behavior (compare Fig.\ 2A and Fig.\ \ref{sigmapfold-null}).
Moreover this deviation becomes bigger as the number of trials $n_t$, in this
case $10$, increases (see Fig.\ 1B).

\end{document}